\documentclass{pasj00}

\begin{document}
\SetRunningHead{K. Matsumoto, T. Kato, and I. Hachisu}%
               {U Sco in the 1999 Outburst: Period Change}
\Received{2000 January 17}
\Accepted{2002 December 11}

\title{The Recurrent Nova U Scorpii in the 1999 Outburst:
       the First Detection of a Significant Orbital-Period Change}

\author{Katsura \textsc{Matsumoto}%
  \thanks{Present address: Graduate School of Natural Science and Technology,
          Okayama University, Okayama 700-8530.}
and
Taichi \textsc{Kato}}
\affil{Department of Astronomy, Graduate School of Science, Kyoto University,
       Kyoto 606-8502}
\email{katsura@kusastro.kyoto-u.ac.jp, tkato@kusastro.kyoto-u.ac.jp}
\and
\author{Izumi {\sc Hachisu}}
\affil{Department of Earth Science and Astronomy, College of Arts and Sciences,
       University of Tokyo, Tokyo 153-8902}
\email{hachisu@balmer.c.u-tokyo.ac.jp}


\maketitle

\begin{abstract}
In this paper we present and discuss our time-resolved photometry
of an eclipsing recurrent nova, U Sco, during an outburst in 1999,
which was conducted from immediately after the optical maximum to
the final fading toward the quiescence.
In the first-ever complete light-curve, a few primary and
secondary eclipses of the binary system were detected,
and the timings of the minima were determined.
We found that the eclipses showed no totality during the outburst.
The depth of the primary eclipses was 0.4--0.8 mag, much shallower
than that in quiescence.
In the plateau phase, very little irradiation ($\leq 0.1$ mag)
was observed in the orbital light curve, which implies the existence
and a large flaring rim of the accretion disk during the outburst.
The minima of the eclipses were detected at earlier orbital phases
for the predicted ephemerides.
Thus, we obtained an orbital period change of the binary
system as $\dot{P}/P = -1.7 (\pm 0.7) \times 10^{-6} ~{\rm yr^{-1}}$
from the $O-C$.
Assuming that this period change is a result of the conservative mass
transfer between the component stars, its mass-transfer rate reaches
$\dot{M} = 2.4 (\pm 1.0) \times 10^{-6} ~M_{\odot}~{\rm yr^{-1}}$
for a $1.37 M_{\odot}$ white dwarf and a $2.0 M_{\odot}$ mass-donor
companion, which is too high to cause shell flashes, even on
a massive white dwarf.
Therefore, this large rate of the period change strongly indicates
a non-conservative mass transfer in the binary system.
\end{abstract}

\section{Introduction}

It is widely accepted that recurrent novae are binary systems
in which mass is transferred to a white-dwarf primary from
a main-sequence or red-giant secondary overflowing its Roche lobe.
The transferred matter gradually accumulates on the surface of
the white dwarf.
When it reaches a critical mass, hydrogen ignites to trigger
a thermonuclear-runaway outburst \citep{starrfield88}
with recurrence periods from one decade to one century.
Recurrent novae are classified as a kind of cataclysmic variables
(e.g., \cite{warner} for a review); at present,
we firmly know only nine systems in our galaxy, i.e.,
CI Aql, V394 CrA, T CrB, IM Nor, RS Oph, T Pyx, U Sco,
V745 Sco, and V3890 Sgr, and one system in the LMC
(e.g., see table 1 of \cite{hachisu01b}).

Although classical novae share many aspects of the outburst nature
with recurrent novae as a subclass of cataclysmic variables,
because the former erupt every ten thousand years or so,
their outburst has been recorded only once in human history.
The white dwarf accumulates hydrogen-rich matter much slower
than that for recurrent novae.
As a result, because a part of the hydrogen diffuses into the white
dwarf before ignition, a very surface layer of the white dwarf
is dredged up into the hydrogen-rich envelope and blown off in
the outburst wind (e.g., \cite{prialnik86}; \cite{kato94}).
Therefore, the ejecta of classical novae contain much of the white-dwarf
matter (carbon, oxygen, or neon), and the white dwarf is gradually eroded.

Recurrent novae, on the other hand, show characteristics clearly
different from classical novae: 1) heavy elements, such as carbon,
oxygen, and neon, are not enriched in ejecta, but are similar to
the solar values (for U Sco, e.g., \cite{barlow}; \cite{williams}),
indicating that the white dwarf is not eroded; 2) very short
recurrence periods from a decade to a century theoretically require
a very massive white dwarf close to the Chandrasekhar mass limit,
$\sim 1.4 M_{\odot}$ (e.g., figure 9 of \cite{nomoto82}).
These two arguments indicate that the mass of the white dwarf
in recurrent novae increases toward the Chandrasekhar mass limit.
We expect that, if the white dwarf is made of carbon and oxygen,
it ignites carbon at the center, and explodes as a type Ia supernova,
or collapses to form a neutron star when it reaches a critical mass
of $1.378 M_{\odot}$ \citep{nomoto84,nomoto91}.

U Sco is a unique member of recurrent novae; among them,
U Sco is characterized by 1) the fastest decline rate of
the light curve in the past outbursts ($\sim 0.6$ mag~d$^{-1}$),
 2) the shortest recurrence period of about 11 yr \citep{schaefer01},
and 3) extremely helium-rich ejecta of He/H $\sim 2$ by number
\citep{barlow,williams}.
The decline rates of the outbursts of recurrent novae are mainly
determined by the white-dwarf masses (e.g., \cite{kato99}),
the rate of which increases as the white-dwarf mass is approaching
the Chandrasekhar mass.
Precise photometry of a recurrent nova in outburst is therefore necessary
to determine the mass of the white dwarf with sufficient accuracy,
especially near the Chandrasekhar limit.
Once the mass is determined, the mass-accretion rate of the white dwarf
can be determined from the recurrence period (e.g., figure 9 of
\cite{nomoto82}).
We are then able to theoretically determine the growth rate of
the white dwarf toward the Chandrasekhar limit.
In the current subject, it is almost certain that U Sco corresponds
to an extreme case of very massive white dwarfs from its fastest
decline rate and shortest recurrence period.
We expect that the recurrent nova U Sco is an immediate progenitor
of a type Ia supernova \citep{hachisu99a}.

Type Ia supernovae are one of the most luminous explosive events of stars.
Recently, they have been used as good distance indicators,
which provide a promising tool for determining cosmological
parameters based on their almost uniform maximum luminosities
(Riess et al.\ 1998; Perlmutter et al.\ 1999).
Both of these authors derived the maximum luminosities ($L_{\rm max}$)
of type Ia supernovae completely empirically from the shapes of
the light curve (LCS) of nearby type Ia supernovae, and assumed
that the same $L_{\rm max}$--LCS relation holds for high red-shift
($z$) type Ia supernovae.
In order for this method to work, the nature of type Ia supernovae and
their progenitors should be the same between high-$z$ and low-$z$ ones.
Therefore, it is necessary to identify the progenitor binary systems,
thus confirming whether or not type Ia supernovae ``evolve'' from
high-$z$ to low-$z$ galaxies (e.g., \cite{umeda}).

It is generally accepted that the exploding star of type Ia supernova,
itself, is a mass-accreting white dwarf in a binary system
(e.g., \cite{nomoto82}; \cite{nomoto84}),
while the companion (and thus the observed binary system) is not known.
Several candidates have been proposed, which include merging double
white dwarfs (e.g., \cite{iben}; \cite{webbink}),
recurrent novae (e.g., \cite{starrfield85}; \cite{hachisu01b}),
symbiotic stars (e.g., \cite{munari92}),
supersoft X-ray sources (e.g., \cite{rappaport}),
etc (e.g., \cite{livio} for more details of these candidate systems).
If U Sco is confirmed to be as massive as or very close to
the critical mass of $M_{\rm Ia} = 1.378 M_{\odot}$, and also
now growing in mass, it is the first strong candidate for
type Ia supernova progenitors.

U Sco is also known as an eclipsing binary system with an ephemeris
for timings of mid-eclipses of HJD 2447717.6061 + 1.23056
$\times E$ \citep{schaefer90,schaefer95}.
The orbital phase duration of the primary eclipses, $\approx 0.1$,
indicates the inclination angle, $i \approx 80^{\circ}$
(e.g., table 4.2 of \cite{warner}).
The origin of binary systems with a very massive white dwarf and
a slightly evolved main-sequence star having a helium-rich envelope
is naturally understood in the context of a binary evolutionary
scenario of type Ia supernovae (\cite{hachisu99a}; \yearcite{hachisu99b}).
Thus, we had been anxiously awaiting the next eruption of U Sco,
since the previous outburst in 1987, in order to obtain a detailed
transition of the nova-outburst, and thus to clarify the physical
nature of the object.

The past outbursts of U Sco were recorded in 1863, 1906, 1936,
1945, 1979, and 1987.
At last, this object underwent its seventh recorded outburst in 1999,
which was detected by \citet{schmeer} on February 25.194(UT)
when the object was about 9.5 mag.
It is noted that the object had been fainter than at least 14.3 mag
on February 25.04(UT) \citep{monard}, and also that there had not been
any positive detection before it \citep{tkato}.
The fast circulation of \citet{schmeer} provided us rare opportunities
for prompt studies of eruption in the very early phase.
Several optical and infrared spectroscopic studies in such a phase
and for later evolution were successfully made by some authors 
\citep{munari99,lepine,anupama,ikeda,evans,thoroughgood}.

We started intensive time-resolved photometry of the outburst
immediately after detection on February 25 in order to understand the
detailed behavior of the progress throughout the event, and then to
probe into its nature.
In the observation, we detected eclipses and revealed that
the orbital period has significantly shortened year by year.
In the following part, details of the observation are described
in section 2, the results are presented in sections 3,
and discussions follow in section 4.

\section{Observation and Data Reduction}

The optical photometric observation of U Sco was carried out
for a total of 30 nights between February 25 and May 20 in 1999,
using an unfiltered CCD camera equiped with a Kodak KAF-0400 chip
(SBIG ST-7) in the 2$\times$2 binning mode, attached to a 0.25-m 
Schmidt--Cassegrain telescope (Meade LX200) installed on
the rooftop of the Department of Astronomy, Kyoto University.
The exposure time for the object was set to 10~s on February 25,
20~s on February 27, and 30~s on the other nights, depending 
on the conditions.
The obtained frames were dark-subtracted, flat-fielded with 
a normalized sky flat, and then analyzed using 
a Java-based aperture and PSF photometry package 
developed by one of the authors (TK).
We performed relative photometry using GSC 6206.372 
[$RA =$ \timeform{16h22m27s.64}, $DEC =$ \timeform{-17D48'57''.5}
(J2000)] as a local comparison star, whose constancy was confirmed
by GSC 6206.266 [$RA =$ \timeform{16h22m45s.79},
$DEC =$ \timeform{-17D51'52''.5} (J2000)].
Although the observational error varied even in an intraday
observation, the typical probable errors for the object were
determined to be $\sim$ 0.1 mag or greater for observations
made on March 2, 5, 9, 11, and 27, which were under poorer
conditions, and on and after March 28 due to the decay of
the outburst; they were determined to be $\sim$ 0.03 mag or better
for the rest of the observations.
The sensitivity of the detector yielded the derived magnitudes
corresponding to those in the $R_{\rm c}$-band.
A heliocentric time-correction was applied for the timing 
prior to analyses.
A journal of the observation is given in table~\ref{tab:obslog}.

\begin{table}
 \caption{Journal of the observation.$^{\ast}$}
 \label{tab:obslog}
 \begin{center}
  \begin{tabular}{lrrcc}
   \hline
   \noalign{\smallskip}
   \multicolumn{2}{c}{Date}   & {$N$} & {Time coverage}   & {Orbital phase} \\
   \multicolumn{2}{c}{(1999)} &       & {(HJD$-$2451000)} & {coverage} \\
   \noalign{\smallskip}
   \hline
   \noalign{\smallskip}
Feb.\ & 25 & 939 & 235.2124--235.3780 & 0.53--0.67 \\
      & 27 & 240 & 237.2038--237.3695 & 0.15--0.29 \\
      & 28 &  91 & 238.2988--238.3578 & 0.04--0.09 \\
Mar.\ &  1 & 303 & 239.2124--239.3712 & 0.78--0.91 \\
      &  2 &  60 & 240.2461--240.3660 & 0.62--0.72 \\
      &  3 & 337 & 241.2275--241.3681 & 0.42--0.54 \\
      &  5 & 167 & 243.1873--243.3662 & 0.01--0.16 \\
      &  9 &   7 & 247.2181--247.3573 & 0.29--0.40 \\
      & 11 &   6 & 249.2529--249.3523 & 0.94--0.02 \\
      & 12 & 435 & 250.1754--250.3622 & 0.69--0.85 \\
      & 13 & 187 & 251.2384--251.3567 & 0.56--0.65 \\
      & 16 & 472 & 254.1679--254.3592 & 0.94--0.09 \\
      & 17 & 410 & 255.1645--255.3359 & 0.75--0.89 \\
      & 22 &  40 & 260.2791--260.3545 & 0.90--0.97 \\
      & 23 & 191 & 261.2753--261.3505 & 0.71--0.77 \\
      & 27 & 109 & 265.3026--265.3446 & 0.99--0.02 \\
      & 28 &  14 & 266.3182--266.3435 & 0.81--0.83 \\
      & 31 & 217 & 269.2315--269.3350 & 0.18--0.26 \\
Apr.\ &  2 & 106 & 271.3025--271.3438 & 0.86--0.90 \\
      &  3 & 174 & 272.2692--272.3373 & 0.65--0.70 \\
      &  4 & 226 & 273.2596--273.3429 & 0.45--0.52 \\
      &  5 &  12 & 274.2452--274.2470 & 0.25--0.25 \\
      &  6 &  63 & 275.3168--275.3402 & 0.12--0.14 \\
      &  8 & 319 & 277.2089--277.3361 & 0.66--0.77 \\
      & 11 & 198 & 280.2595--280.3357 & 0.14--0.22 \\
      & 16 & 102 & 285.3011--285.3384 & 0.24--0.27 \\
      & 29 &  72 & 298.1201--298.1415 & 0.65--0.67 \\
      & 30 &  55 & 299.1138--299.1344 & 0.46--0.48 \\
May   &  1 & 100 & 300.0980--300.1365 & 0.26--0.29 \\
      & 20 &  15 & 319.0657--319.0708 & 0.68--0.68 \\
   \noalign{\smallskip}
  \hline
  \end{tabular}
 \end{center}
 $^{\ast}$ $N$ is the number of data points used in the analyses.
 The orbital phase was calculated using the ephemeris given in
 Schaefer and Ringwald (\yearcite{schaefer95}).
\end{table}

\section{Result}

\subsection{Light Curve of the 1999 Outburst}

\begin{figure}
 \begin{center}
  \FigureFile(88mm,61.6mm){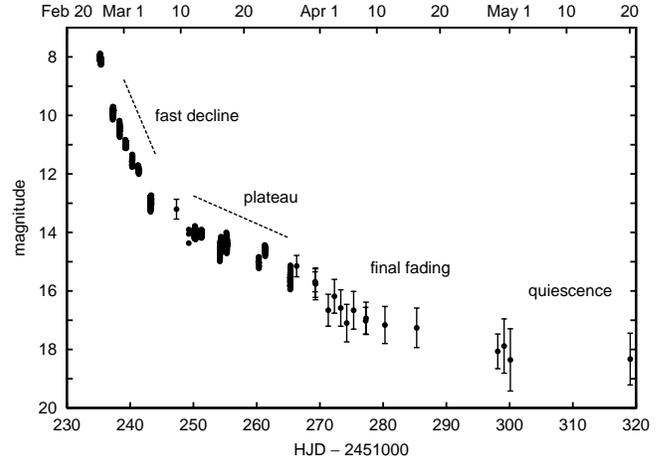}
 \end{center}
 \caption{Optical light curve of the 1999 outburst of U Sco.
 The vertical axis represents the magnitude corresponding to
 the $R_{\rm c}$ magnitude, and the lower and upper horizontal axes
 represent HJD$-$2451000 and the date in 1999, respectively.}
 \label{fig:lc_all}
\end{figure}

In the observation we obtained the first-ever complete light-curve
of U Sco in outburst covered from immediately after the optical
maximum to the final decay toward quiescence through the
mid-plateau phase, which is shown in figure~\ref{fig:lc_all}.
We found that the overall evolution was consistent with the results
presented in \citet{munari99} and \citet{kiyota},
 and that this outburst closely followed the light curve made with
the past outbursts superposed (e.g., \cite{rosino}).

We divided the whole light curve of the present outburst
into three stages of 1) fast decline phase, 2) mid-plateau phase,
and 3) final fading phase toward quiescence, according to
the decline rates (figure~\ref{fig:lc_all}).
In the fast decline stage, the object showed an extremely rapid
decline of $\sim 0.59$ mag~d$^{-1}$ until March 13, i.e., about
16 days after the optical maximum.
Such a fast decay is known as to be one of the characteristic features
seen in outbursts of U Sco.
The light curve was momentarily flatten to around 14 mag on
March 11--13 at the end of this first stage.
The declining rate then turned to be gradual ($\sim 0.095$ mag~d$^{-1}$)
between March 16 and 28 (plateau phase).
The object finally entered into the last decline stage
with a declining rate of $\sim 0.19$ mag~d$^{-1}$
about 32~days after the maximum (final fading phase).
At that time we could not perform time-resolved photometry
due to the faintness of the object, and therefore the data points
were averaged each night from the start of this third stage
(and also on March 9 due to an extremely bad sky condition),
as shown in figure~\ref{fig:lc_all}.
After the final fading phase, the object returned to quiescence around 
the end of 1999 April.

\subsection{Primary and Secondary Eclipses}

\begin{figure}
 \begin{center}
  \FigureFile(88mm,61.6mm){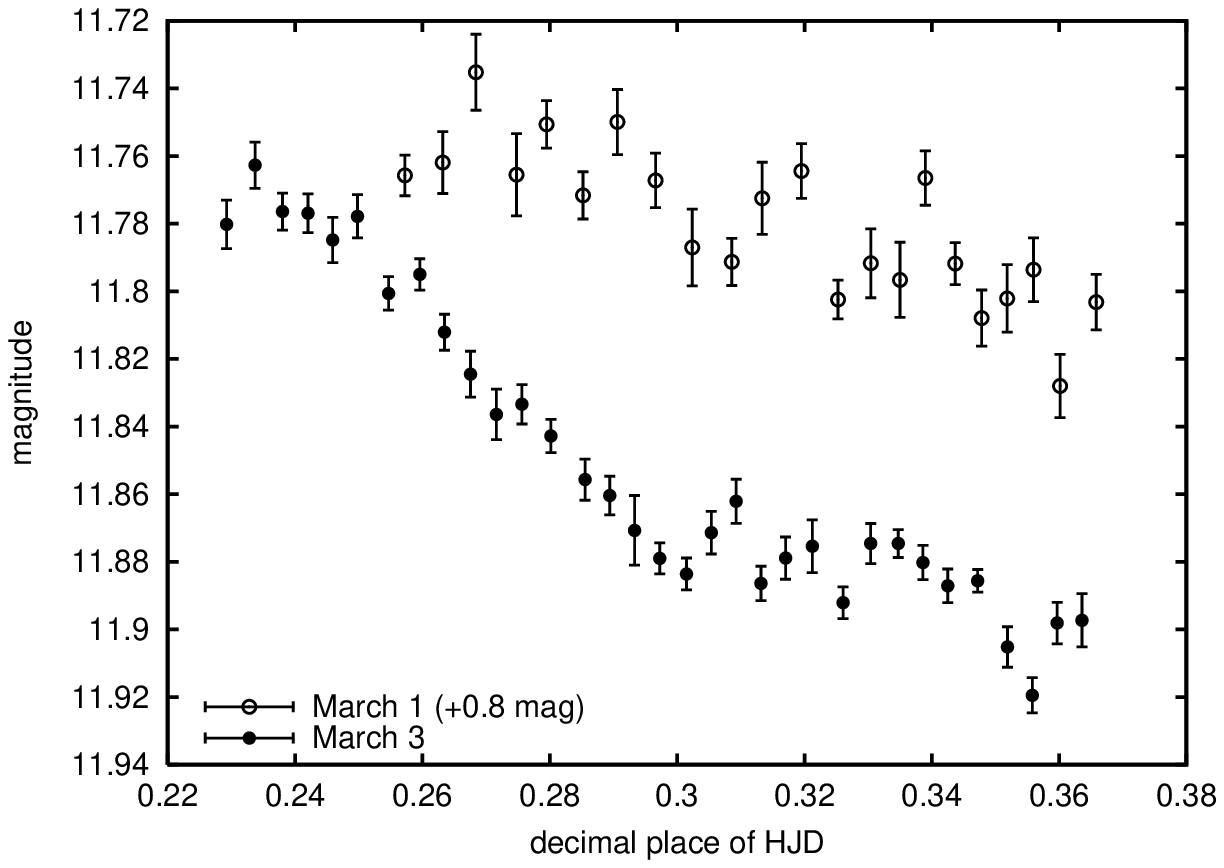}
  \FigureFile(88mm,61.6mm){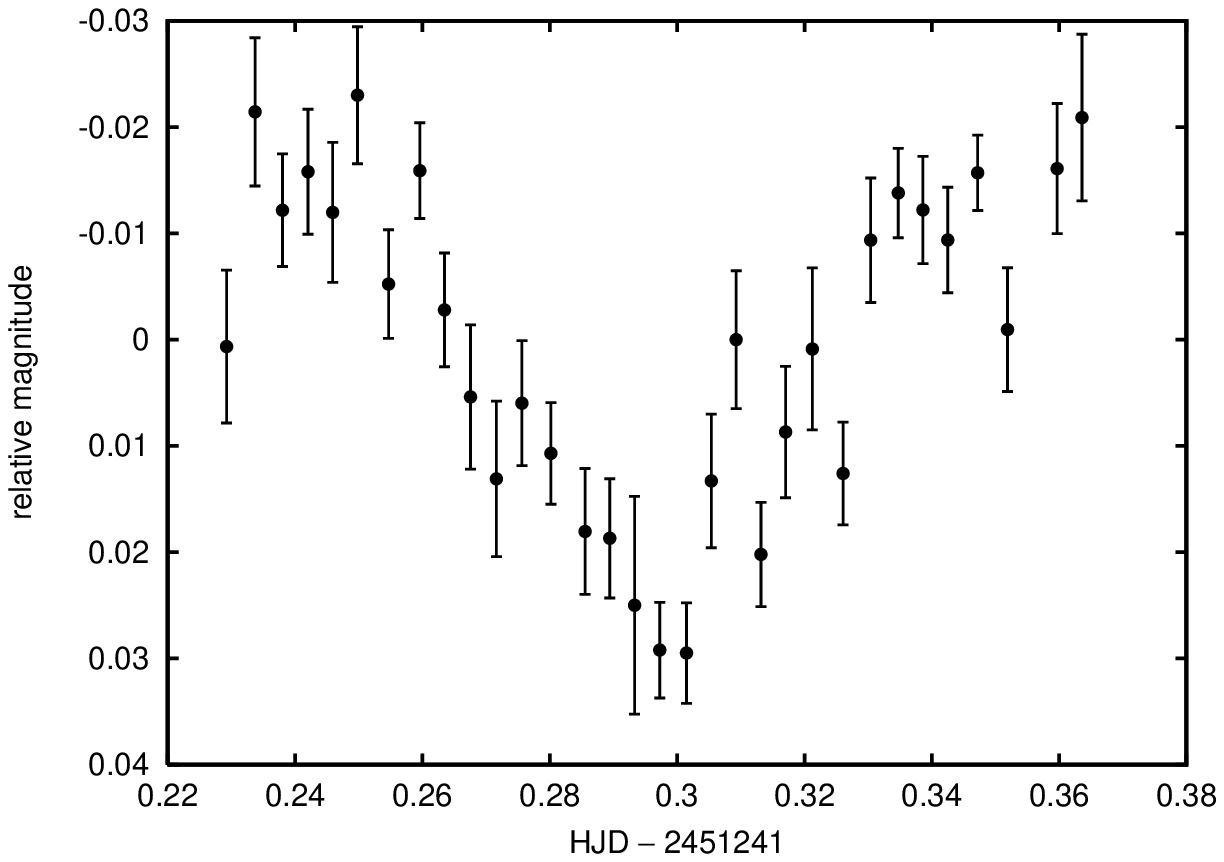}
 \end{center}
 \caption{[Upper panel] Secondary eclipse of U Sco observed on 
 March 3 (filled circle).
 The light curve on March 1 from which relatively noisy points were
 partially excluded (open circle)
 is also presented in order to illustrate the significance of
 the detection of the eclipse on March 3.
 In the light curves each data point was averaged by each 10 successive
 points for clarity, and each point is provided with estimated errors.
 [Lower panel] The light curve on March 3 after removing the local
 linear trend in order to demonstrate the eclipsing feature.}
 \label{fig:lc_0303}
\end{figure}

In the early part of the fast decline phase, we obtained almost
linear declining trends on each night, especially on February 25
and March 1 with longer coverages with continuous $\sim 0.13$ orbital
phase durations and better qualities with typical probable errors
of $\leq$ 0.01 mag.
In the fast decline phase, however, we detected a secondary
eclipse of the binary system in the observation on March 3.
The light curve on that night showed a clear intraday modulation
digressing from a linear declining trend, and no similar
features were seen in the other nightly light curves during
this phase (figure~\ref{fig:lc_0303}).
Note that the linear trend which involves all affectable
factors for the observation, such as extinction, is not identical
to the fast declining trend of the outburst denoted by the dashed
line in figure~\ref{fig:lc_all}.
It should also be noted that the curvature for the local light curve
is extremely small, i.e., the quadratic term around 10 days is about
$-0.03$ mag~d$^{-2}$, indicating that the influence of the curvature,
about 0.0014 mag for the 0.2 d duration, should be safely negligible.
We confirmed no significant change in the counts of the comparison
star, GSC 6206.372, other than the quite normal slow variation
caused by changing airmass throughout the entire run.
The differential photometry using the check star, GSC 6206.266,
also yielded no significant variation larger than the observational
error.

The detrended light curve on March 3 revealed that the minimum
occurred at HJD 2451241.298 along with the depth of
the eclipse being about 0.05 mag, and its duration was at least
0.1~d, i.e., 0.08 orbital phase.
This detection was only one week after the optical maximum,
which is the earliest example of eclipses among those after
optical maxima in the past nova eruptions have ever been observed.

The secondary eclipse observed is theoretically expected
in a numerical model of \citet{hachisu00a}
as being the consequence of an irradiation effect by the expanded
photosphere of the white dwarf in the nova outburst.
In the fast decline phase, the photosphere of the white dwarf
is as large as, or larger than, the size of the binary system.
On March 3, because its photospheric radius was still as large as
$1.8 R_{\odot}$ (the binary separation of $\sim 6 R_{\odot}$),
the irradiation of the companion star was large enough to
make a secondary eclipse with a depth of 0.1 mag, from a theoretical
model of a thermonuclear runaway event.

\begin{figure}
 \begin{center}
  \FigureFile(88mm,61.6mm){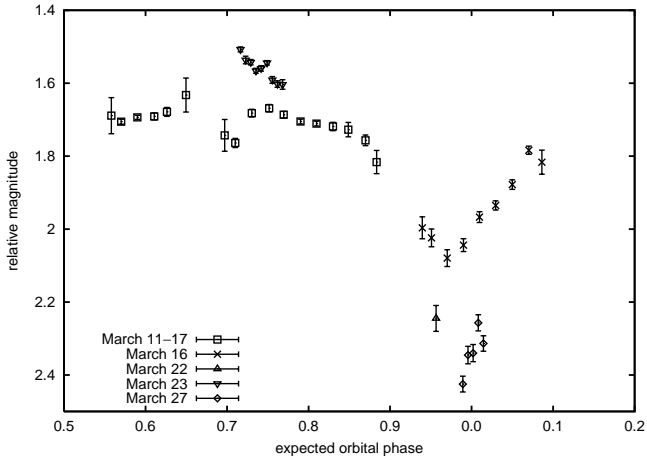}
 \end{center}
 \caption{Orbital light curve of U Sco during the plateau phase
 of the 1999 outburst, which was constructed from the fragmentary
 light curves on March 11--13, 16, 17, 22, 23, and 27.
 Each data point was averaged by each of about 60 points on March 11--17
 and each of about 20 points on March 22--27 for clarity.
 The data on March 16 and 27 corresponding to the primary eclipses
 are separately presented from the compiled light curve.
 The horizontal axis represents the orbital phases based on
 the orbital period and the ephemeris of the minima of the primary eclipse
 of U Sco defined by Schaefer and Ringwald (\yearcite{schaefer95}).
 The contrast between the data on March 22 and 23 is reasonably 
 explained by the eclipsing feature on the each occasion based on 
 the ephemeris.
 The vertical axis represents the relative magnitudes.
 The declining trend during the plateau phase was removed prior to 
 combining each light curve.}
 \label{fig:lc_0316}
\end{figure}

In addition to the secondary eclipse, we detected a minimum and
ensuing egress of a primary eclipse of the binary system at
HJD 2451254.2 in the observation on March 16; also, an ingress
of the next primary eclipse was observed after
the 1.23~d orbital period on the next night.
In figure~\ref{fig:lc_0316} the primary eclipse observed on March 16
corresponding to an orbital phase coverage of 0.9--0.1 is shown.
The timing of the minimum occurred at HJD 2451254.211.
The expected orbital phase given in the light curve in
figure~\ref{fig:lc_0316} is based on the ephemeris determined
by Schaefer and Ringwald (\yearcite{schaefer95}),
and the data points are averaged by
each phase bin of 0.02 for clarity, after removing the internight
linear decline of 0.095 mag~d$^{-1}$ for the plateau phase.
The light curves on March 11--17, 22, 23, and 27 are also presented
in order to depict the features outside the primary eclipse.
Note that only data points outside of eclipse are shown for
the data on March 11 in figure~\ref{fig:lc_0316} because of
the poor sky condition.
Since no other nights fall on a range of 0.9--0.1 orbital phase 
during the interval of March 11--17, the profile of the eclipse 
minimum on March 16 is not influenced by the detrending process 
among the different nights.
The orbital phase between 0.1 and 0.5 was unfortunately not covered 
during this plateau phase.

The object was observed as being somewhat below the declining
trend-line on a few nights.
The light curves on March 5 and 22 are such cases,
as can be seen from figure~\ref{fig:lc_all}.
The scatter of the data on March 5 was mainly caused by larger
observational errors, and no systematic variation was observed
in the light curve.
However, the orbital phase corresponded to 0.01--0.15, which is
consistent with the fact that the object really showed a primary
eclipse at that time.
Because the observed duration on March 22 corresponded to an orbital
phase of 0.90--0.97, it is also probable that the light curve could be
related to a primary eclipse, and the observation on March 23
corresponded to an orbital phase of around 0.75.
Hence, the light curves are consistent with the binary phases and 
the difference of the averaged magnitudes between March 22 and 23 
is therefore reasonably explained. 
Nevertheless, it would be better that we reserve our conclusion about
whether these two sets really display the primary eclipses or not,
and therefore we have not given a clear discussion for these nights.
We also briefly mention that the number of data points on
March 28 was small due to the poor sky condition on that night;
we thus conclude that the data are not sufficient to discuss
the intraday variabilities during that night.

\subsection{Orbital Variability during the Outburst}

The light curve shown in figure~\ref{fig:lc_0316} demonstrates
the shape of the primary eclipse of U Sco in an outburst.
The primary eclipse on March 16 had a depth of about 0.4 mag
in the early plateau phase of the outburst, which is much shallower
than $\Delta\,B$ $\sim$ 1.5 mag determined in quiescence
\citep{schaefer90,schaefer95}.
The light curve also showed no totality of the minimum and,
accordingly, indicates that the primary eclipse was a partial
occultation.
In figure~\ref{fig:lc_0316} the data on March 27 are also
presented as an example of the shapes of the primary eclipses
during the later plateau phase of the outburst.
This additional data set revealed that the primary eclipse of U Sco
was deeper and steeper in the later plateau phase.
We conclude that the amplitudes of the primary eclipses are
0.4--0.8 mag in the plateau phase, and that we were observing
a growth of the eclipsing depth as the outburst was fading away.
The data during May 11--17 suggest that the duration of the eclipse
extended to 0.3--0.4 orbital phase, which is wider than the $\sim$ 0.2
orbital phase determined in quiescence. 

The orbital light curve also provided us the physical states of
the binary system during the outburst.
An extremely small reflection effect ($\le 0.1$ mag) is
suggested from the light curve during the orbital phase of
$\sim$ 0.55--0.9, which indicates that irradiation of the companion
star by the white dwarf is shielded during the plateau phase.
Thus, we can reasonably conclude that the accretion disk should 
have been remaining after the nova explosion (see also \cite{sekiguchi})
and that the rim of the disk was largely flaring-up in the plateau phase
to shield the light from the white dwarf.

We divided the whole light curve of the outburst into
three stages in subsection 3.1.
The secondary eclipse detected on March 3 occurred in the first
decline phase, during which the photosphere of the white dwarf was
still expanded up to $1.8 R_{\odot}$ which was comparable to
the size of the binary system \citep{hachisu00a}.
As a result, the companion (mass-donor star) could be strongly
irradiated by the white-dwarf photosphere.
In the plateau phase, however, the white-dwarf photosphere
shrunk to $0.1 R_{\odot}$, or smaller.
Therefore, it could be blocked by the flaring-up rim of the
accretion disk.
In the final decay phase, the white-dwarf photosphere returned to
its almost original radius in quiescence.

It is noted that no significant variation, like flickering reported
by \citet{schaefer90}, was observed in the light curves of the individual
nights.

\section{Discussion}

\subsection{Non-Conservative Mass Transfer and the Orbital-Period Change}

\begin{figure}
 \begin{center}
  \FigureFile(88mm,61.6mm){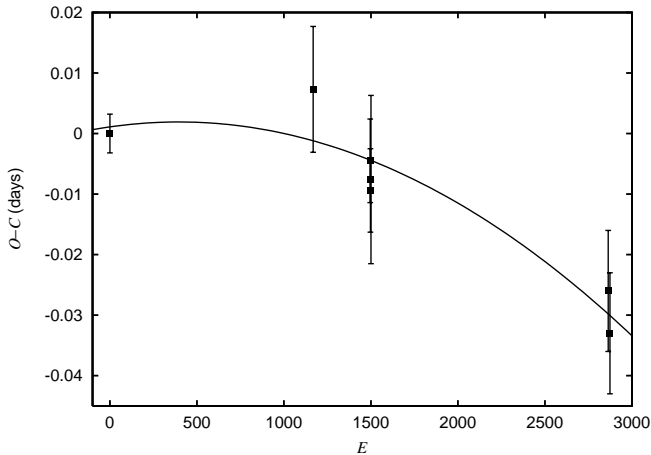}
 \end{center}
 \caption{$O-C$ diagram derived from the timings of the eclipse
 minima based on the data in Schaefer and Ringwald (\yearcite{schaefer95})
 and this work.
 The horizontal and vertical axes represent the cycle number $E$
 and $O-C$ value in days, respectively.
 The solid line indicates the obtained period change rate of 
 $\dot{P}/P = -1.7 \times 10^{-6}$ yr$^{-1}$.}
 \label{fig:o-c}
\end{figure}

The timing of the minimum observed in the secondary eclipse
on March 3 was centered at HJD 2451241.298, while the predicted
ephemeris was HJD 2451241.324 based on the orbital ephemeris of
Schaefer and Ringwald (\yearcite{schaefer95}).
In addition, a significantly advanced timing of the eclipse was
also detected in the primary eclipse observed on March 16.
The center occurred at HJD 2451254.211, which was 0.027 orbital
phase, i.e., 0.033~d, earlier than the expected minimum of
HJD 2451254.244 based on Schaefer and Ringwald's ephemeris.
The error in these timings was estimated to be 0.01~d,
by applying the method of \citet{fernie} on the mean error
to Lafler--Kinman class of statistics,
along with the consistency of the multiple detections of the eclipses
including the secondary one on March 3.
Hence, the latter result strongly confirmed the negative $O-C$
(observed minus calculated timings) inferred from the secondary eclipse;
moreover, these two timings at the minima agree remarkably well with
each other in the $O-C$ diagram (figure~\ref{fig:o-c}).

The consistency is significant; it supports the correctness of
the determinations of the eclipse minima, which statistically
excludes the possibility of an accidental coincidence of such
timings in $O-C$.
We may consider other possible variabilities: for example,
\citet{budzinovskaya} reported their observation of U Sco
in the 1987 outburst, and claimed the existence of a short-time
modulation with a 0.053~d periodicity, while they did not deny
the possibility that their variation might be the orbital-period
modulations given by \citet{schaefer90}.
However, the 0.053~d periodicity cannot be applied to our cases,
because it is too short and deep to explain our minima lasting
for at least 0.1 orbital phase.
The possibility of an apparent period change observed in a fading object
is known in dwarf-nova outbursts \citep{krzeminski}.
It is safely excluded for the present case because of the non-total
eclipse shapes and the dominance of the disk in the optical range
instead of a hotspot usually observed in dwarf novae.
\citet{thoroughgood} have argued that the minimum which occurred at
HJD 2451286.2 was consistent with the ephemeris of
Schaefer and Ringwald (\yearcite{schaefer95}),
but their sampling in the photometry was not sufficiently accurate to
detect the minimum.
The direct use of the individual minima in table~1 of
Schaefer and Ringwald (\yearcite{schaefer95})
might have introduced a large uncertainty in the calculation of the
$O-C$ fit.
According to the error values of the timings, it is more secure
that the minima in each season are treated as representatives
for the three seasons in evaluations of $O-C$.

We therefore conclude not only that the minima detected in our
observation were surely consequences of eclipses, but also that
the timings of the minima had advanced relative to the expected ones.
These shifts of the timings imply that the orbital period of
the binary system had significantly shortened, likely due to
mass transfer, during the last quiescent state.
Accordingly,
the orbital-period change of the binary system obtained
by a least-squares fitting among the available data of
Schaefer and Ringwald (\yearcite{schaefer95}) and
the present result is derived as
\begin{equation}
 \frac{\dot{P}}{P} = -1.7 (\pm 0.7) \times 10^{-6} ~{\rm yr^{-1}},
\end{equation}
where $P$ is the orbital period.

If this period change is the result of a conservative mass transfer
between the components of the binary system,
the mass-transfer rate is given by
\begin{equation}
 \dot{M_{2}} = - \frac{\dot{P}}{P} \frac{M_{2}}{3 (1 - q)},
\end{equation}
where $M_{2}$ is the mass of the secondary star and $q \equiv M_{2}/M_{1}$.
This yields
\begin{equation}
 \dot{M_{2}} = - 2.4 (\pm 1.0) \times 10^{-6} ~M_{\odot}~{\rm yr^{-1}}
\end{equation}
for a white dwarf with $M_{1} = 1.37 M_{\odot}$ and 
a slightly evolved main-sequence star with $M_{2} = 2.0 M_{\odot}$.
However, this mass-accretion rate is too high to cause shell flashes,
and it results in steady hydrogen-shell burning on very massive
white dwarfs instead (e.g., figure~9 of \cite{nomoto82}).

\citet{hachisu00a} estimated the envelope mass
at the optical maximum as being $\sim 3 \times 10^{-6} ~M_{\odot}$.
If this is the present case, the mass-accretion rate should be
smaller than $\sim 2.5 \times 10^{-7} ~M_{\odot}~{\rm yr^{-1}}$
in the quiescence between 1987 and 1999, because a part of
the helium-layer mass may have been dredged up
(e.g., \cite{kovetz}; \cite{prialnik95}).
The mass-transfer rate of $\sim 2.4 \times 10^{-6} ~M_{\odot}~{\rm yr^{-1}}$
derived from equation~(3) above, which resulted in an envelope
mass of $\ge 10^{-5} ~M_{\odot}$ during the previous quiescence,
is again too high to be compatible with the envelope
mass derived by \citet{hachisu00a}.

Hence, we are forced to consider the possibility of non-conservative
mass transfer in the U Sco system.

Assuming that a part of the transferred matter was lost from the binary
system through the outer Lagrangian points, \citet{hachisu00b} derived
a mass-transfer rate of 
$\dot{M_{2}} \sim -5.5 \times 10^{-7} ~M_{\odot}~{\rm yr^{-1}}$
for a 0.8--2.0 $M_{\odot}$ mass-donor star during the previous quiescence.
This implies that a part of the transferred mass, i.e.,
$\sim 3.0 \times 10^{-7} ~M_{\odot}~{\rm yr^{-1}}$, was being lost
from the binary system.
The outflowing matter carried away the orbital angular momentum
effectively, and caused a shrink of the orbital separation.

\subsection{U Sco as a Progenitor of Type Ia Supernovae}

Together with the ephemeris of Schaefer and Ringwald (\yearcite{schaefer95}),
we detected a significant change in the orbital period for the first time.
A brief summary of our theoretical analysis for the present
study was published in \citet{hachisu00a}.
Their main results are summarized as follows:
1) The mass of the white dwarf was determined as 
$M_{1} = 1.37 \pm 0.01 M_{\odot}$ by fitting their model
light curves with the early 10 days of optical data.
Because the fast decline rate of the light curve during the early
10 days is very sensitive to the mass of the white dwarf, but
almost independent of the mass of the companion, the configuration
of the accretion disk, or the hydrogen content of the envelope,
this determined mass is rather stiff.
2) The envelope mass at the optical maximum was
$\sim 3 \times 10^{-6} ~M_{\odot}$ and
3) the hydrogen content of the envelope was
$X = 0.05$ by assuming a solar metallicity of $Z = 0.02$.
4) Therefore, the mass-accretion rate of the white dwarf was
$\sim 2.5 \times 10^{-7} ~M_{\odot}$~yr$^{-1}$ during the
quiescent phase between 1987 and 1999, if no white-dwarf matter
was dredged up into the envelope.
5) About 60\,\% ($\sim 1.8 \times 10^{-6} ~M_{\odot}$) of the
envelope was blown off in the outburst wind, but 40\,\%
($\sim 1.2 \times 10^{-6} ~M_{\odot}$) was left and added
to the helium layer on the white dwarf.
6) As a result, the white dwarf can grow in mass at a rate of 
$\sim 1.0 \times 10^{-7} ~M_{\odot}$~yr$^{-1}$.
Thus, they concluded that U Sco is a strong candidate
for a progenitor system of the type Ia supernova, because the white 
dwarf is very close to the critical mass for type Ia supernova 
explosion, and will be able to reach the critical mass of 
$M_{\rm Ia} = 1.378 M_{\odot}$ long before the mass-donor is exhausted. 

\citet{thoroughgood} independently estimated the mass of
the white dwarf by means of radial velocity studies of the optical
spectra obtained in the final fading phase of the 1999 outburst.
The mass which they determined, based on the mass function of the binary
system, was extremely close to the Chandrasekhar limit
($M_{1} = 1.55 \pm 0.24 M_{\odot}$); they concluded that U Sco is
the best candidate binary system growing to a type Ia supernova.
Their result supports our conclusion about the nature of U Sco,
described above, except for the slightly lighter estimation of
the mass of the secondary star.

\subsection{Relation to Supersoft X-Ray Sources}

In the 1999 outburst of U Sco, the object was observed as a supersoft
X-ray source by the BeppoSAX satellite on March 16--17, which was
about 19 days after the optical maximum \citep{kahabka}.
It was an observational confirmation of a theoretical prediction
based on the thermonuclear runaway of recurrent novae \citep{kato96},
and the first definite case of recurrent novae, implying a similarity
to supersoft X-ray sources.
We have recently come to know a second possible case, the eclipsing
recurrent nova CI Aql, which underwent the second recorded
eruption in 2000 \citep{kiss,matsumoto,lederle,matsumoto2}.
Although a firm confirmation has still not been made,
CI Aql is expected as to be a recurrent nova related to supersoft
X-ray emissivity, according to a theoretical model for
the light curve of the 2000 outburst of that object
\citep{hachisu01a,hachisu03}.

Supersoft X-ray sources are characterized by luminous and
extremely soft X-ray emissions (e.g., \cite{sss} for a review).
Part of those sources are plausibly accreting binary systems
containing a relatively massive white dwarf with
nuclear burning of accreting matter on the surface
\citep{vdh,rappaport}; several systems have been identified
in our Galaxy and the Magellanic clouds at present \citep{sss_cat}.
We discuss here similarities of U Sco to supersoft X-ray sources.

CAL 87 is a typical example of supersoft X-ray sources
to understand the photometric characteristics of U Sco,
since CAL 87 is also an eclipsing binary system
(e.g., \cite{hutchings} and references therein).
A light-curve analysis for CAL 87 has suggested a flared
accretion-disk rim and irradiations of the disk and secondary star
\citep{schandl}.
In the case of U Sco, the primary eclipse which we observed on March 16
just corresponds to that in the supersoft X-ray state.
Taking into account both a flaring-up effect of the accretion-disk
rim and an expansion of the accretion-disk radius, which exceeds
the Roche lobe size, \citet{hachisu00a} successfully reproduced
the shallow depth of the primary eclipse (figure~3 of \cite{hachisu00a}).

Recently, several systems of cataclysmic variables have also been
identified as supersoft X-ray sources, which are the so-called V Sge-type
stars \citep{steiner,patterson,greiner98L} and VY Scl-type stars
\citep{greiner98,greiner99}.
An observationally interesting and remarkable fact concerning those
systems is that supersoft X-rays are commonly detected only during
faint states in optical wavelength.
Such a fact was first observed in the supersoft X-ray source
RX J0513.9$-$6951 \citep{reinsch,alcock,southwell}.
The same matter is also suggested for other members of supersoft 
X-ray sources, showing such switches of bright/faint states
in relatively long-term light curves, such as RX J0019.8+2156 
\citep{greiner95,simon} and RX J0527.8$-$6954 \citep{greiner96}.

These features indicate a self-absorption effect of supersoft X-rays,
which only appears during optically bright states.
In the case of U Sco, an optically thick wind should cease before
the detection of supersoft X-rays in the plateau phase.
This feature can be understood by a massive wind produced by
the white dwarf during the optically bright state, i.e., the early phase
of the outburst.
\citet{hachisu00a} have shown that the strong wind stops 17 days
after the optical maximum, and that the duration of the optically
thick wind is consistent with the supersoft X-ray detection.

We detected the growth of the eclipse depth, from about 0.4 mag
to 0.8 mag, in the plateau phase of the outburst; interestingly,
these depths in the bursting phase are much shallower than
$\sim 1.5$ mag in $B$-magnitude obtained in quiescence \citep{schaefer90}.
It is well known that the eclipse in bright states of
V Sge is much shallower than that in the faint states
(e.g., \cite{herbig}; \cite{patterson}).
The present detection of a new example of the shallow primary
eclipse implies that, although its physical mechanism is not yet
clear, such a shallower depth is not special in the 1999 outburst
of U Sco, but common in supersoft X-ray sources when they are
in a bright state or in a bursting state.

\section{Summary}

We obtained a complete light curve of the 1999 outburst of U Sco,
including a few detections of eclipses of the binary system.
The decline of the outburst showed three trends, and the light
curve generally traced those of the past recorded ones.
The most substantial result in this study was the detection of
a significant advance of the timings of the eclipse minima,
which should be a consequence of an orbital-period change
of the binary system during the previous quiescence.
We established the orbital-period change for the first time;
the mass-transfer rate derived from the observed $\dot{P}/P$
strongly indicates the occurrence of a non-conservative mass transfer
in U Sco.
The behavior of the orbital light curve observed in the outburst 
suggests a common configuration in the optical high state of
the supersoft X-ray sources (including transient sources).

\bigskip

The authors would like to express their sincere gratitude to the VSNET
members, especially to Patrick Schmeer, who was involved in the
detection and confirmation of the outburst of this object in 1999.
KM was financially supported by a Research Fellowship of
the Japan Society for the Promotion of Science for Young Scientists.
This work was partly supported by a Grant-in-Aid for
Scientific Research (10740095, 11640226, 13640239) of the
Ministry of Education, Culture, Sports, Science and Technology.


\end{document}